\documentclass{article}
\usepackage{amsmath,amssymb}
\usepackage{epsfig}
\date{}

\begin{document}
\title{MYRON MATHISSON:\\
 WHAT LITTLE WE KNOW OF HIS LIFE
 \thanks{Based on talks given by the authors at the conference
 ``Myron Mathisson: his life, work, and influence on
 current research'', Stefan Banach International Mathematical Center,
  Warszawa, 18--20 October 2007}}

\author{Tilman Sauer\\
\small{Einstein Papers Project, California Institute of Technology 20-7}\\
\small{Pasadena, CA 91125, USA}\\[0.2cm]
        Andrzej Trautman\\
\small{Institute of Theoretical Physics, University of Warsaw}\\
\small{Ho\.za 69, 00-681  Warszawa, Poland}}

\maketitle

\begin{abstract}
Myron Mathisson (1897-1940) was a Polish Jew known 
for his work on the equations of motion of bodies in general relativity
and for developing a new method to analyze the properties of fundamental solutions
of linear hyperbolic differential equations. In particular, he derived
the equations for a spinning body  moving in a gravitational  field
and proved, in a special case, the Hadamard conjecture on the class of
equations that satisfy the Huygens principle. His work still exerts
influence on current research.
Drawing on various archival and secondary sources, in particular his correspondence
with Einstein, we outline Mathisson's biography and scientific career.

\medskip

\noindent
PACS numbers: 01.65.+g, 01.30.Cc
\end{abstract}

\section*{Early years}

\begin{figure}[htb]
\begin{center}
\includegraphics[scale=0.4]{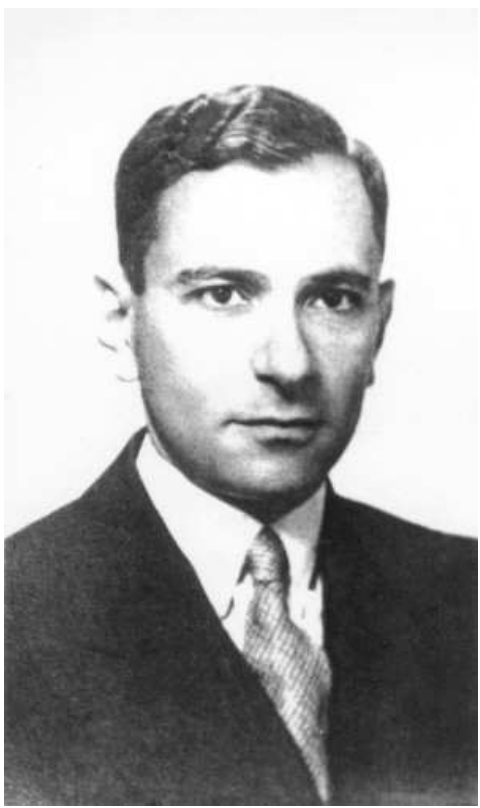}
\caption{Myron Mathisson (1897--1940)}
\end{center}
\end{figure}

Myron Mathisson was born in Warsaw on 14 December 1897.%
\footnote{Unless stated otherwise, biographical information about
    Mathisson's early life is taken from a letter to Einstein, dated
    23 February 1930 (AEA 18-004). See the Appendix for a synopsis of the available
    sources for biographical information about Mathisson and a resolution of
    our abbreviations such as (AEA) and (UW). On his high school diploma (UW), issued on
    1 May 1915, the date of his birth is given, in the old-style (Julian) calendar,
    as 2~December 1897. This corresponds to 14 December 1897 in the
    Gregorian calendar. In the twentieth century, the difference between those
    calendars was 13 days; this may have led to the fact that some sources, 
    including AEA 18-004, give December 15 as the relevant date.}
Very little is known about his family background. His parents,
Hirsh and Khana Mathisson, had moved from Riga to Warsaw (IG). We do not know
the profession or business of Myron's father, but the
family must have been relatively well-off, as may be inferred from
their address in a well-to-do area of Warsaw (Polna 70, now Noakowskiego 16),
as well as from the excellent education received by Mathisson.
He attended, from 1906 to 1915, a Russian philological
gymnasium named after {\it The Great Prince Aleksy Nikolayevich, The
Successor to the Throne\/}, one of the best secondary schools in
Warsaw at the time.
His high school diploma contains grades in Russian, Latin, Greek,
German and French (Polish was then not allowed in schools), but also in
mathematics and physics; all his grades were excellent, and he graduated
with a gold~medal.

We do not know what language was spoken in the Mathisson household;
Myron knew perfectly well Hebrew, Russian, and Polish; he never mentioned Yiddish.

\section*{1915-31 Studies and Ph.D.}

On 3 December 1915, Mathisson started studies at the Department of Civil Engineering of
Politechnika Warszawska, which at that time was already a
Polish language school. (Recall that, as a result of the war, the Russians
were forced, by the German army, to leave Warsaw).  He explained later in a letter to Albert
Einstein that his decision to study at this school had been influenced
by the fact that excellent French mathematicians and physicists of the
times of Augustin Fresnel had studied at the \'Ecole
Polytechnique. One can guess that he soon realized
that the Warsaw Polytechnic was not quite like the French one:
in 1917 he started to follow the activity of the
physics laboratory at Warsaw University.

On 11 June 1917, Mathisson received a severe reprimand from the Senate
of the Warsaw Polytechnic for ``having signed an indecent and improper
polemic with the Rector and Senate of Politechnika Warszawska.'' (PW) It is
unknown, however, what the ``polemic'' had been about.  After passing
the First Diploma examination in 1918, he quit the Polytechnic and,
in 1919, was drafted into the Polish army, then fighting the Soviets.

After the war,  he tried, unsuccessfully, to resume studies at
the Warsaw Politechnic. In the Archives of that school, there is
a very polite letter by Mathisson, dated 6 October 1920, addressed to the Rector,
asking for readmission. The official who dealt with the case noted on the margin
that Mathisson had left the Politechnic after having received a reprimand and submitted
his case to the Senate; this led to a rejection (PW).

In the end the refusal turned out good
for science because, as a result, in the fall of 1920, Mathisson
entered Warsaw University to study physics. However, the death of his
father, some time in the early 1920s, resulted in a deterioration
 of the financial situation of
the family and forced him to interrupt his
studies.

In 1925 Mathisson wrote his first scientific paper {\it On the motion of a
rotating body in a gravitational field}.
Czes{\l}aw Bia{\l}obrzeski, the only professor of theoretical
physics at Warsaw University at the time, was ready to accept it as
a Ph.~D.\ thesis, but Mathisson became dissatisfied with his own work, left
the tutelage of Bia{\l}obrzeski without even saying a word of
good bye, as he later admitted to Einstein, and went to Palestine.
It is unclear how long Mathisson stayed in
Palestine, but a severe tropical illness, probably malaria, forced him to
return to Poland.

After having returned to Warsaw, he continued to work, in solitude, on
the problem of motion in the general theory of relativity.
His small income was derived from
giving Hebrew lessons and performing some auxiliary tasks for
the construction engineers.%

\begin{figure}[htb]
\begin{center}
\epsfig{file=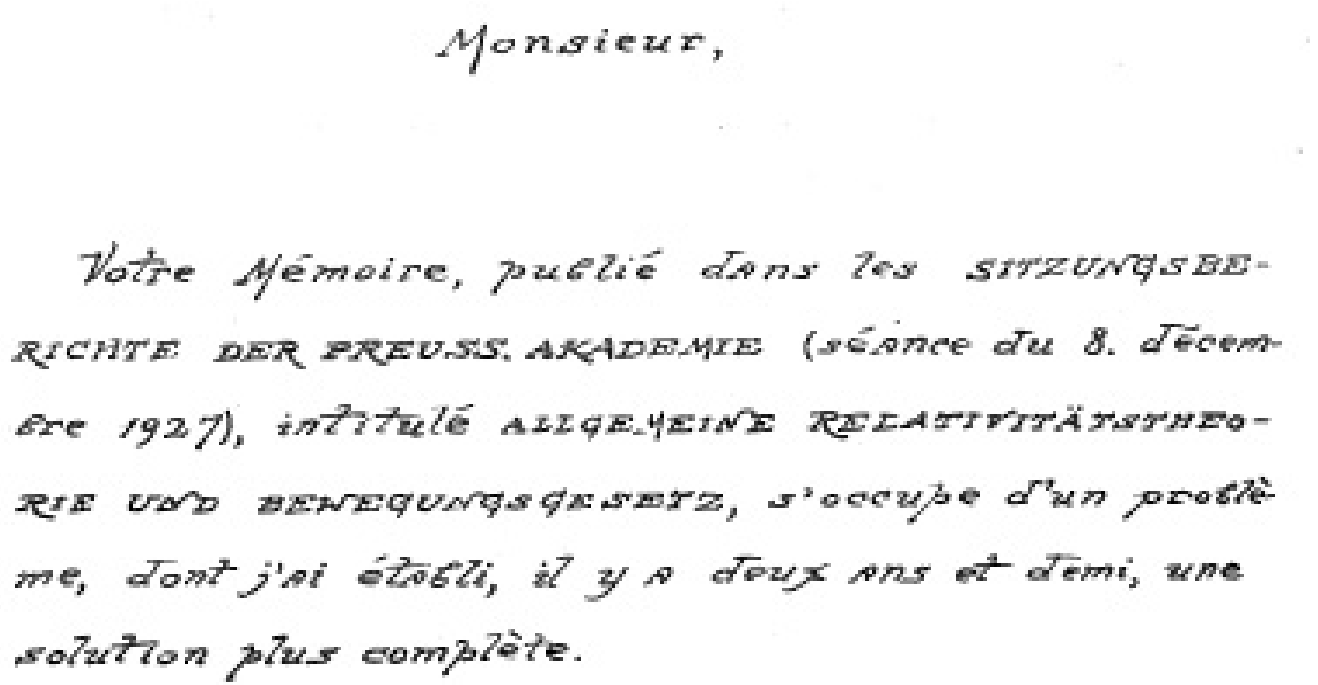,width=11.5cm}
\caption{The beginning of Mathisson's first letter to Einstein, 18 December 1929
(AEA 18-001).}
\label{Fig2}
\end{center}
\end{figure}

On 18 December 1929, Mathisson started his correspondence with Albert Einstein%
\footnote{The Albert Einstein Archives (AEA) contains 19 letters by Mathisson
    to Einstein, and 11 letters by Einstein to Mathisson. In
    addition, the archives contain an unpublished manuscript by
    Mathisson (AEA 18-030) and more than 30 letters that mention
    Mathisson. All of Mathisson's correspondence is hand-written,
    while all of Einstein's extant letters in the archives are
    preserved as carbon copies of typewritten letters. It is
    clear, both from gaps in the sequence of the correspondence
    and from reference to missing letters, that not all of
    Einstein's letters have been preserved.  Presumably, it is the
    handwritten letters by Einstein that we are missing.
        \label{noteAEA}}
with a long first letter (see Fig.~\ref{Fig2}).%
On reading it, one is impressed by the quality of his French
and the audacity of his criticism of Einstein's paper: Mathisson
writes that, in \cite{Ein27}, Einstein neglects radiation and deviations
from spherical symmetry, these approximations being due to
the ``mathematical insufficiency of your method''.

At the time of writing this letter, the 32-year-old Mathisson did not
have a Ph.D.\ nor any publications to his reputation. Yet, in this
11-page letter Mathisson described, in general terms, without
equations, his views on the problem of motion in general relativity,
boldly criticizing Einstein's own approach. He also gave an indication
of his difficult financial situation. He announced that he would send
Einstein a manuscript detailing his ideas, but said that, should this
manuscript be published, he would have to ask for some financial
remuneration for it.

Einstein's reply to Mathisson's first letter is unknown, but from Mathisson's
second letter, dated 14 February 1930 (AEA 18-002),
we can infer that Einstein had replied obligingly by inviting Mathisson to come
to Berlin to collaborate with him in some form.
Einstein at the time was, of course, a world-famous
man. Scientifically, he was deeply involved in investigating
the implications of his teleparallel approach toward unified
field theory, and he was, in fact, just looking for another
mathematically trained collaborator
\cite{Sauer2006}.  His long-time assistant Jakob Grommer had just left
or was just about to leave to take up a position in Minsk. Cornelius (Korn\'el)
L\'anczos, who had spent a year with Einstein in Berlin to work with him
on the teleparallel approach, was leaving as
well. In early 1930, Walther Mayer came to Berlin and became
Einstein's mathematical collaborator for the next few years. We should
also like to mention that Einstein himself had written similarly
audacious letters to scientific authorities such as
Paul Drude, when he had still been a young man, unknown to the
scientific world.  Mathisson declined the invitation on the ground
that he did not yet feel prepared for such a collaboration. But he did
send a manuscript on the problem of motion.

An exchange of several letters followed; in his response of 20
February 1930 (AEA 18-003), Einstein wrote that a superficial reading
of Mathisson's paper had already convinced him ``that you have an
extraordinary formal talent and that you are made for scientific
work''.  He suggested that Mathisson's manuscript be submitted
as a Ph.D.\ thesis. He offered to try to obtain a fellowship 
that would enable Mathisson to come to Berlin and work with
him. To this end, he asked Mathisson for further information about his
life, academic training, and profession.  In response,
Mathisson sent a third letter, dated 23 February 1930 (AEA 18-004), 
in which he gave a curriculum vitae of sorts,
another outline of his ideas and comments on his personal and intellectual
situation. This letter of 14 pages is one of the most
important sources for our knowledge about Mathisson's early life and work.

In response to Mathisson's third letter, Einstein forwarded Mathisson's
manu\-script to Bia{\l}obrzeski and suggested that Mathisson be awarded
a Ph.D. degree on the basis of his results on the problem of motion,
apologizing for Mathisson's earlier ``odd behavior''. He also offered to
cover the expenses connected with presenting the thesis (AEA 18-005, 18-006).
Despite his enthusiasm for Mathisson's work, Einstein also asked whether
``it might be possible to make do without Schouten's terrible notation?
This would be such a relief for the reading mankind''. The notation in question
concerned integrals over \(n\)-dimensional manifolds. Mathisson did not follow
Einstein's suggestion and, in later letters (AEA 18-027 and  18-034), explained
how this precise notation helped him to obtain new results.

On 31 October 1930 M.~Mathison%
\footnote{This spelling was in all legal documents, such
as his passport, high school certificate, and student identity card;
in his letters and publications, he always spelled his name  with a double s,
a logical thing to do, considering the Latin form (Matthaeu{\bf s}) of the name
Matthew.}
obtained the Ph.D. degree from Warsaw University on the basis of a
thesis on {\it General relativity theory and the dynamics of the
electron\/} (in Polish) (UW).

\begin{figure}[tbh]
\begin{center}
\includegraphics[scale=0.55]{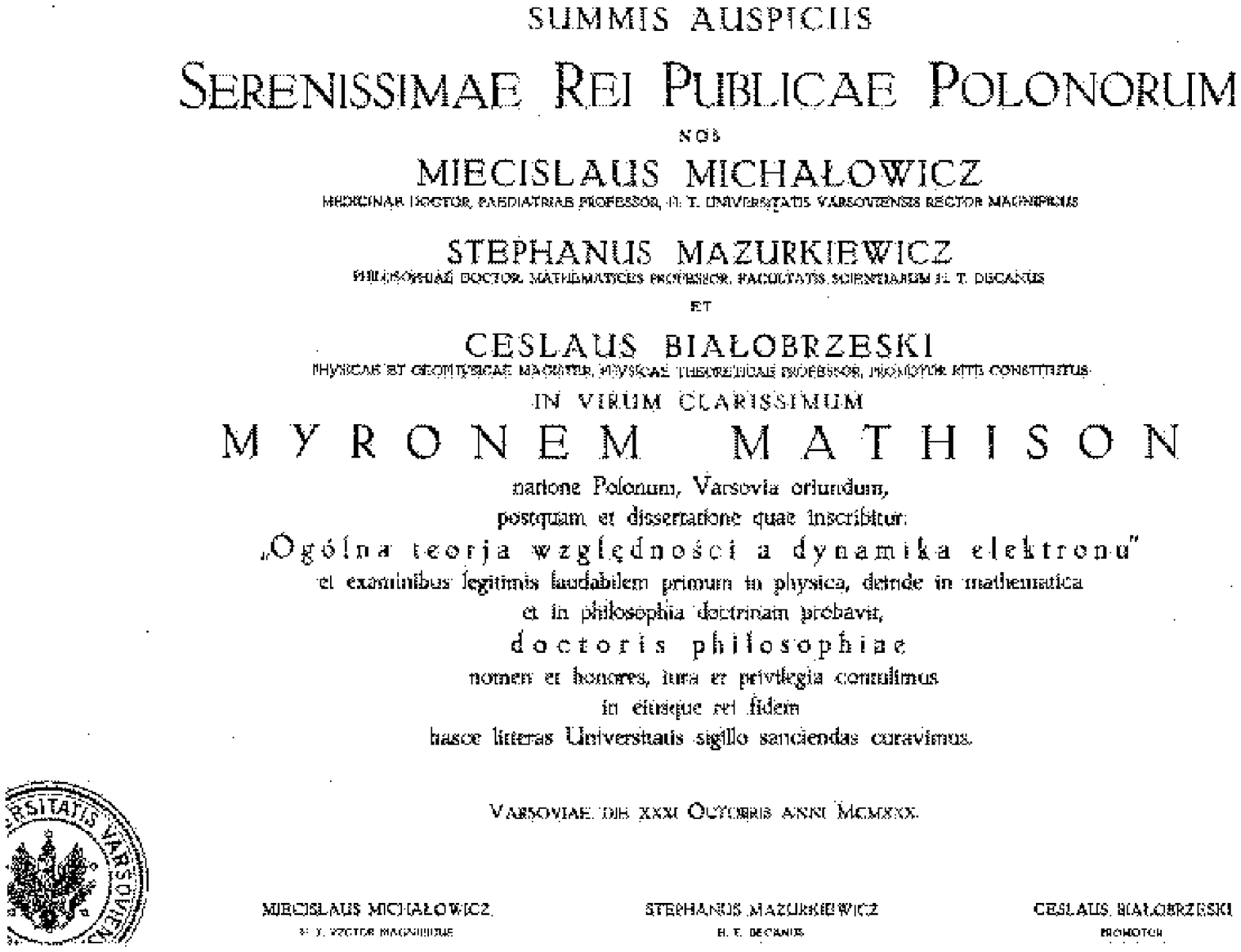}
\caption{Myron Mathisson's doctoral diploma  (UW). Note the spelling of his name.}
\end{center}
\end{figure}

The text of the thesis has not been
preserved, but its results are presented in Mathisson's first three
publications.  These appeared in 1931 in {\it Zeitschrift f\"ur
Physik} [M1-3]%
\footnote{In this style we refer to the list of {\it
Scientific publications of Myron Mathisson\/} included in the last
section of this paper.}
with Einstein's help.  A week after obtaining
his Ph.D., Mathisson wrote to Einstein informing him of his
graduation as well as about the Rockefeller Foundation's negative decision
(see below; AEA 18-029). He
expressed his hope to obtain the scholarship if he succeeded in
publishing his results. He indicated that he would like to come to Berlin
to see Einstein but also because that would bring him ``nearer to his
final destination---Palestine''. He also included a manuscript, a
``small physical sketch'' as he called it, entitled ``Relativistic
Formulation of Quantum Phenomena'' (AEA 18-030).

In his response (AEA 18-031), Einstein rejected Mathisson's ideas on the quantum problems
on the ground that his approach would be equivalent to the old, non-relativistic quantum theory
based on the Bohr--Sommerfeld quantization rule \(\int pdq = nh\).
 In response to a follow-up letter by Mathisson, in which
he indicated his intention to publish his relativistic results of the last
three years (AEA 18-032), Einstein sent Mathisson two letters of recommendation ``on the
basis of which your relativistic works, that are certainly very valuable, will
be accepted with certainty'' (AEA 18-033).

\section*{The first publications}

Mathisson's first three papers indeed appeared in {\it Zeitschrift f\"ur Physik}. [M1]
was received by that journal on 8 December 1930, [M2] on 23 December 1930, and
[M3] on 19 March 1931.

Those papers contain an implicit polemic with Einstein and his
approach to the problem of motion. The first paper contains an essential generalization of
Einstein's linearization of the field equations \cite{Ein16}.
Mathisson allows the background metric  to be curved.
 In the second, Mathisson shows that
nonlinearity of the field equations is not essential for obtaining from
them the equations of motion. He uses extensively the geometry of the null
cone introduced earlier by Hermann Minkowski \cite{Mink}.
In  flat  space-time, given a time-like world-line \(z^\mu(s)\),%
\footnote{We use the  Greek alphabet to denote space-time indices. This is in
 agreement with the prevalent current notation, but Mathisson, in his first papers,
  used  Latin indices. The index \(\nu\) after a comma (resp., semicolon) denotes
  the ordinary (resp., covariant) derivative in the direction of the vector
  \(\partial/\partial x^\nu\).}
he  introduces two functions of \(x=(x^\mu)\): a null coordinate \(s\)  and a co-moving
radial distance \(r\),
such that the vector \(l^\mu(x)=x^\mu-z^\mu(s(x))\) is null and
oriented towards the future and \(r=\dot{z}_\mu l^\mu \). If the
electromagnetic  potential is assumed to have the form
\(A^\mu=e^\mu(s)/r\) so that it satisfies \(\Box A^\mu=0\), then the
Lorentz condition \({A^\mu}_{,\mu}=0\) implies
\(e^\mu(s)=e(s)\dot{z}^\mu(s) \) and the conservation of charge,
\(\dot{e}=0\). Therefore, the Li\'enard--Wiechert potential
\(A^\mu=e \dot{z}^\mu /r\) satisfies Maxwell's equations for an
arbitrary motion of the charge. After describing this, Mathisson
[M2] then considers an analogous problem for a weak gravitational
field \(h_{\mu\nu}\). He introduces, following Einstein, the tensor
\(\psi_{\mu}^{\nu}=h_{\mu}^{\nu}-\tfrac{1}{2} \delta_\mu^\nu
h^\rho_\rho \) so that the linearized field equations in empty space
are equivalent to \(\Box \psi_{\mu}^{\nu}=0\) and
\({\psi^{\nu}_{\mu,\nu}}=0\).  The tensor field
\(\psi^{\mu\nu}=m^{\mu\nu}(s)/r\) satisfies the wave equation and
the  Einstein condition (so called in [M3])
  \({\psi^{\nu}_{\mu,\nu}}=0\)
implies \(m^{\mu\nu}=m\dot{z}^\mu\dot{z}^\nu\), \(\dot{m}=0\) and
the equation of motion \(\ddot{z}^\mu=0\). In the third paper, he
mentions the possibility of using Dirac's delta functions, a novelty
for relativists at that time. He gives a new derivation of the
special relativistic equation of motion of a charge, taking into
account its self-interaction, resulting in the radiative reaction
force \(\tfrac 2 3 e^2 (\ddot{u}^\nu u_\nu u^\mu- \ddot{u}^\mu)\),
where \(u^\mu=\dot{z}^\mu\).

Early in 1931, Mathisson wrote his first paper on a new approach to the study of
fundamental solutions of partial differential equations of the
hyperbolic type; in mid-March, he sent a manuscript, entitled ``A parametrix method
for generalized wave equations in Riemannian manifolds'' to Einstein. With a little delay
due to his being absent from Berlin, the paper was forwarded by Einstein to
Otto Blumenthal, managing editor of {\it Mathematische Annalen}. Einstein, who had
been co-editor of the journal from 1919 to 1928, recommended the paper
for publication, saying
\begin{quote}
I know the author from earlier works as a very intelligent and diligent writer
who is dealing with the deepest problems of the general theory of relativity under
very difficult external conditions. (AEA 18-036)
\end{quote}
In a postcard written a day later, Blumenthal acknowledged receipt of the manuscript,
but requested that Einstein make more specific comments on the manuscript itself. No paper
with that title ever appeared in the {\it Mathematische Annalen}, and we
may conjecture that it appeared only later with a similar title in 1934 in the Polish
journal {\it Prace Matematyczno-Fizyczne} [M5].
Nevertheless, a paper by Mathisson on the same subject, but with different title [M4],
appeared later in the {\it Mathematische Annalen}. This paper, however, was only received
by the journal on 21 December 1931, and its publication was probably endorsed by
Jacques Hadamard. In this and in later papers on this subject, Mathisson made essential use
of methods of differential geometry, tensor calculus, and the geometry of the null
cone. In his letter of 25 April 1931 (AEA 18-034), he emphasized the essential role
that  tensor calculus and Riemannian geometry
 had played in his  derivation of an integral formula in the paper
that he was sending then to Einstein.

\section*{Efforts to obtain a Rockefeller stipend for Mathisson}

In 1930, after reading Mathisson's first letters and manuscript,
Einstein  sent several letters to various officers of the {\it
Notgemeinschaft der Deutschen Wissenschaft} and of the Rockefeller
Foundation to inquire about the proper procedure of applying for a
stipend for Mathisson. As it turned out, the rules required that the
application be filed not by Einstein but by a mentor of the
candidate's home country, Bia{\l}obrzeski in Mathisson's case.%
\footnote{For a historical account of the role of scholarships awarded by the
    International Educational Board of the Rockefeller Foundation
    for the internationalization of mathematics in the twenties of the
    last century, see \cite{Sieg}.}
More importantly, several conditions had to be satisfied. The candidate had
to present an invitation by a scholar abroad whose guest he would be,
he had to hold ``at least'' a Ph.D.\ and have scientific publications
under his name, and he had to have an academic position in his home
country to which he would be able to return after spending time
abroad.  Only the first condition was satisfied in Mathisson's case. So
Einstein wrote another diplomatic letter to Bia{\l}obrzeski (AEA 18-009), asking
him to submit the application on Mathisson's behalf, and, in
particular, to be generous in his assurances of a future academic
career for Mathisson:
\begin{quote}
    It will hardly be avoidable, for the sake of a
    good cause, to pay tribute, albeit somewhat insincerely, to the
    almighty Bureaucratius, which a higher authority
    will readily forgive because of its unreasonableness. (AEA 18-009)
\end{quote}
Bia{\l}obrzeski complied with Einstein's wish (AEA 18-010), and Einstein, on his
part, sent a letter of recommendation in support of Mathisson to the
Rockefeller Foundation (AEA 18-011).

In June 1930, a Rockefeller delegate met with Mathisson, but missed
Bia{\l}obrzeski. Informed by Bia{\l}obrzeski about the failed meeting,
Einstein wrote another letter to the Rockefeller Foundation, confirming his
support of Mathisson's application. In spite of Einstein's intervention,
the Rockefeller Foundation informed Bia{\l}obrzeski that its Paris office
would withhold the application and not forward it to the Committee of the
International Educational Board in the U.S.\ until Mathisson had fulfilled the
necessary requirements.

During his visit in the United States, on 14 December 1930,
Einstein had a chance to meet John D.~Rockefeller and his wife in person (RA).
The meeting was arranged on Einstein's initiative
and during their conversation, Einstein tried to convince Rockefeller that his rules
for awarding scholarships were too strict and often defeated the very purpose of the
original intent of the scholarship program. In an account of the meeting by a mutual
acquaintance, George Sylvester Viereck, who had arranged it, we read
\begin{quote}
Professor Einstein argued with him that the strict regulations laid down by his
educational foundations sometimes stifled the man of genius. ``Red tape,'' the
professor exclaimed, ``encases the spirit like the bands of a mummy.''
Rockefeller, on the other hand, pointed out  the necessity for carefully
guarding the funds of the foundations from diversion to unworthy ends or
individuals who are not most meritorious. Standing his ground against the
greatest mind in the modern world, he ably defended the system under which
the various foundations were conducted. ``I,'' Einstein said, ``put faith
in intuition.'' ``I,'' Rockefeller replied, ''put my faith in organization.''
Einstein pleaded for the exceptional man. Rockefeller championed the greatest
good of the greatest possible number. Einstein was the aristocrat, Rockefeller
the democrat. Each was sincere, each without convincing the other, persuaded him
of his sincerity. \cite{Viereck}
\end{quote}
Mathisson, as it turned out, would become a victim of this conflict of opinions.
The immediate outcome of Einstein's visit to Rockefeller was a promise that
Rockefeller himself would look into any particular case that Einstein might point out
to him where the rules of his foundation exclude a worthy recipient.
After his return to Berlin in spring 1931, Einstein wrote to Mathisson that he wanted
to make another attempt to obtain a Rockefeller scholarship for him (AEA 18-035).
Now that Mathisson had obtained his Ph.D.\ and had also a couple of scientific publications,
most of the earlier obstacles had been removed. The only condition that was still
violated was the requirement that Mathisson be given assurances of a permanent academic
position in Poland to return to after his stay abroad. Once again, an application
for a scholarship was formally filed but again it seems to have never been forwarded
from the European office in Paris to the awarding committee in the U.S. In June
1931, Einstein even wrote to Rockefeller himself, mentioning Mathisson as a case
in point of their earlier dispute (AEA 18-049). In a polite response, Rockefeller
promised to look into this but indicated that he could do so only after the summer
when the committee would meet again (AEA 18-050).
This was very much the end of Einstein's attempt
to obtain a Rockefeller scholarship for Mathisson.

In September 1931, Mathisson wrote to Einstein (AEA 18-046), in a somewhat depressed mood,
that the Rockefeller Foundation once more had decided not to forward his application,
that he had saved some money and could visit Einstein in Berlin;
he also mentioned that he was considering looking for a job in
Russia or in Palestine, because
 {\it ``in Polen
ist kein Platz da f\"ur Leute meiner Nationalit\"at}''. The visit was
not realized because Einstein was then about to go to the United
States. He wrote back that one should first wait what would come of his
intervention with Rockefeller, and then see about this after his return to
Germany (AEA 18-048). As it turned out, neither did Mathisson ever receive a scholarship
from the Rockefeller foundation, nor did Einstein and Mathisson ever meet.

\section*{1932--35: Habilitation, lectures in Warsaw and
first visit to Paris}

In 1932, Mathisson obtained a habilitation at Warsaw University that
allowed him to use the title of ``docent'' (analog of the German {\it
Privat-Dozent\/}) and give lectures (already during the year
1932--33), but did not imply a permanent position (UW).

During the  years 1932--36, Mathisson gave at Warsaw University, in
his capacity of docent, several courses of lectures: {\it Kinetic theory
of gases\/}, {\it Applications
of the theory of groups to quantum theory\/} (based on { Wigner}'s
book), {\it Theory of relativity,} {\it Tensor calculus\/}
(according to Schouten),   {\it Cosmology} and {\it Theoretical Physics\/}
for students of chemistry. He also conducted, together with
Bia{\l}obrzeski and Otton  Nikodym, the main seminar on theoretical
physics (UW and HU).

Mathisson's work on partial differential equations attracted the attention
of  Hadamard who, in 1933, wrote to Einstein saying that
his recommendation could secure a stipend for Mathisson to come to
Paris (AEA 12-040). Einstein, then himself a refugee in Belgium, wrote such a
letter of recommendation for Mathisson to the mathematician Paul Montel
in Paris (AEA 18-051). Mathisson continued to work on partial differential equations and
published, in 1934, in a Polish journal, another paper on the subject [M5].

In 1935, at the invitation of Jacques Hadamard and Paul Langevin,
 Mathisson went to Paris and gave lectures at the Coll\`ege de France
 on differential equations of the hyperbolic type and the diffusion
 of waves in Riemannian spaces (HU).

\section*{1936-37: appointment in Kazan}

On 3 November 1935 Einstein wrote to Hadamard saying that Mathisson could come
for a year to the Institute for Advanced Study in Princeton and asking for
his current address (AEA 18-053).
He also wrote a letter to Mathisson, sent to Paris and containing
an invitation for Mathisson  (no copy in the Archive). The letters were
forwarded by Mlle  Jacqueline Hadamard, a daughter of the
mathematician, to Mathisson who was then in Russia (as were many scientists during the
1930's: Jakob Grommer, Nathan Rosen, Victor Weisskopf, and others).

Presumably, the letters took a long time to reach Mathisson; his reply
was dated ``Moscou, le 23 juin 1936'' (AEA 18-054). Delighted by the invitation, Mathisson
wrote that he could come to Princeton for the academic year 1937--38,
after a year spent in Moscow and Kazan. He also commented favorably
on the conditions of work in Kazan and implied that he would go back
to Kazan from Princeton.

From a letter of Prof.~B.L.~Laptev we know that Mathisson arrived in
Kazan on May 3, 1936 (BLL), but from Mathisson's letter
it follows that he was spending a part of the summer of that year
in Moscow.

We do not know how and by whom Mathisson's appointment in Kazan
had been arranged.

In a letter of 7 July 1936, Einstein expressed delight with the news
that Mathisson had found good conditions
to work in Soviet Russia (AEA 18-055). He wrote that,
considering how many scientists had then been deprived of the possibility to
work, it would not be right to invite Mathisson to the Institute for
Advanced Study. In a postscript, Einstein mentions that he has shown, with Rosen, that
there are no gravitational waves. There is an excellent
account by Daniel Kennefick \cite{Ken05} of the events that occurred
in connection
with an attempt by Einstein to publish a paper with Rosen purporting to
show the non-existence of plane gravitational waves.

Mathisson's next letter, from Kazan, is dated 18 April 1937 (AEA 18-057); he
refers to Einstein's results on gravitational waves
as being in agreement with his work on Huygens' principle and the diffusion
of waves in curved spaces. He expects to travel in June
to Paris  and gives, for correspondence, his mother's address in Warsaw
(Leszno 47, not as good as the previous one).

A letter of 7 May 1937 (AEA 18-058), is the last, in the Einstein Archives, from
Einstein to Mathisson. It is a little cooler than earlier letters;
Einstein wrote about his collaboration with ``your colleague'' Leopold Infeld
(and with Banesh Hoffmann). They have developed a new method (later called
the EIH method) of
deriving the equations of motion of point masses and have shown
that there are no additional conditions
that could be interpreted as corresponding to quantum phenomena \cite{EIH}.
By May 1937, arrangements had already been made with Leopold
Infeld to come to the Institute for Advanced Study in order to work
with Einstein. Infeld had asked about the possibility to do so
in February 1936, in early May his salary for a one-year stay at the
Institute had been granted, and Infeld eventually arrived in Princeton
in October 1937 \cite{Sta}.

Mathisson left Kazan at the end of May 1937, and went on a short visit to
Paris; in a letter sent to Einstein on 5 September 1937 (AEA 18-061), he explained
that he would not return to Kazan because ``already at the end of May
the situation of a foreigner there was unbearable''; he left behind
all his belongings and books (presumably so as not to let  the
Soviet authorities know of his intentions). Recall that in 1937
Stalin's purges and spy-hunting were intensifying.%

A few months after returning from Kazan, Mathisson sent to {\it Acta
Physica Polonica\/} his most important  paper [M6].
In this paper, Mathisson introduced
the notion of a ``gravitational skeleton'' and gave a derivation of
the coupling between spin and curvature.  In Mathisson's definition of the
gravitational skeleton one can see the germ of the idea of a
distribution, as later introduced by Laurent Schwartz:
Mathisson uses ``test functions'' \(p_{\mu\nu}\) and uses the equation
\begin{equation}\label{V}
    \int_{D}T^{\mu\nu}p_{\mu\nu}d^4x=\int_{L}(m^{\mu\nu}p_{\mu\nu}+
m^{\mu\nu\rho}p_{\mu\nu;\rho}+\dots)ds
\end{equation}
to replace the continuous energy-momentum tensor \(T^{\mu\nu}\) filling
a world-tube \(D\)
by an equivalent -- as far as the external field is concerned -- distribution
with support on a time-like world-line \(L\subset D\).
Taking  \(p_{\mu\nu}=\xi_{\mu;\nu}+\xi_{\nu;\mu}\)
and using the conservation law \({T^{\mu\nu}}_{;\nu}=0\),
Mathisson proves that the above integral over \(L\) vanishes
for arbitrary \(\xi\) vanishing at both ends of the world-line
(``Mathisson's variational principle''). He then uses this principle to
derive the equations of motion, which in modern form are \cite{Dixon70}
\begin{eqnarray}\label{e:MP}
\dot{p}_\mu+\tfrac 1 2 R_{\mu\nu\rho\sigma}u^\nu s^{\rho\sigma} &=&0,\\
\label{e:MP2}
\dot{s}_{\mu\nu}+u_\mu p_\nu-u_\nu p_\mu &=&0,
\end{eqnarray}
where  \(p^\mu\) and \(s^{\rho\sigma}\) are, respectively, the momentum and
spin (intrinsic angular momentum) of a body
moving in space-time with curvature described by the Riemann tensor \(R_{\mu\nu\rho\sigma}\);
dots denote covariant derivatives in the direction of \(u\).
They were derived also by Achilles Papapetrou \cite{Papa} and are now often
called the {\it Mathisson--Papapetrou equations}.
In the rest frame of the body, spin is presumed to have only space components;
this is expressed by the Frenkel condition \cite{Fr26},
\begin{equation}\label{e:F}
s_{\rho\sigma}u^\sigma=0,
\end{equation}
which leads to \(p_\mu=mu_\mu+s_{\mu\nu}\dot{u}^\nu\), where \(m=p_\mu u^\mu\).
Equations \eqref{e:MP} and \eqref{e:MP2} subject to \eqref{e:F},
in the limit of special relativity, have
``helical'' solutions that Mathisson interpreted as being classical
counterparts of the quantum {\it Zitterbewegung\/} [M7]. Tulczyjew \cite{Tul59}
proposed to replace \eqref{e:F} by the condition \(s_{\rho\sigma}p^\sigma=0\)
which, in special relativity, together with \eqref{e:MP} and \eqref{e:MP2},
 implies that the center-of-mass line
\(L\) is straight, \(p_\mu=mu_\mu\)  and \(\dot{s}_{\mu\nu}=0\).
The equations \eqref{e:MP} and \eqref{e:MP2} have been the subject of much
research and were thoroughly discussed at the conference; they were also derived
for a Weyssenhoff--Raabe fluid with spin, within the framework of the Einstein--Cartan theory
\cite{Traut72}.

In September 1937 Mathisson attended a conference at the Institute for Theoretical
Physics (now Niels Bohr Institute) in Copenhagen (NBA).

\begin{figure}[htb]
\begin{center}
\includegraphics[width=9cm]{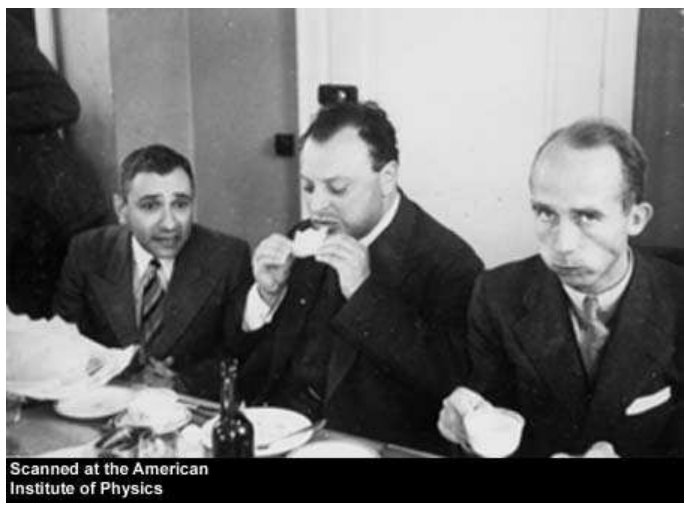}
\caption{Mathisson, Wolfgang Pauli and Gerhard Heinrich Dieke in Copenhagen
 (1937). Photo: AIP.}
\end{center}
\end{figure}
\section*{Candidacy for the chair for theoretical physics in Jerusalem }

For some time, Mathisson was being considered for a chair for theoretical
physics at the Hebrew University of Jerusalem. The search for this position
went on for a number of years, and it was difficult to find a suitable
candidate for the task \cite{Unna}. Mathisson's name seems to appear for
the first time in a letter by Adolf Abraham Ha'levi Fraenkel, then dean of the faculty
of sciences of the Hebrew University, to Einstein of 18 June 1936. After Eugene Wigner's
candidacy had fallen through, Fraenkel mentioned Mathisson as a candidate together
with Infeld and also Louis Goldstein of Paris. In his response, Einstein
proposed Lothar Nordheim and Cornelius L\'anczos over Mathisson and Goldstein.

In 1937, Einstein's opinions about Mathisson with respect to the position in
Jerusalem were curiously intertwined with their mutual work on the problem of
motion.
On 18 April 1937, Mathisson wrote to Einstein: ``I am happy to tell you that
I have solved the quantum problem as a dynamical problem of the general
theory of relativity...'' (AEA 18-057). Ten days later, Einstein recommended
Infeld over Mathisson for the position in Jerusalem in a letter to Salman (Shlomo)
Schocken, member of the governing board of the university
(AEA 37-251). A week later, on 7 May 1937, Einstein responded to Mathisson's
earlier letter, saying: ``your indications remain completely incomprehensible to
me...'' (AEA 18-058). Several weeks later, on 16 June 1937, the EIH paper \cite{EIH}
was submitted
to {\it Annals of Mathematics}. On 4 July 1937, Mathisson wrote to Einstein again
announcing ``exact results'' and asking him to forward these to {\it Physical Review}.
Just a few days earlier, on 1 July 1937, Fraenkel had written to Einstein that
they are considering Infeld ``even though we have received new recommendations about
Mathisson'' (AEA 37-254). Fraenkel probably referred to a letter, written in Hebrew, dated
``Warsaw, 13 June 1937,'' and signed $\aleph$ (aleph), which highly praised Mathisson
for the position in Jerusalem (HU). In response to Fraenkel's letter, Einstein wrote on 
22 July 1937 to Fraenkel,
confirming his assessment that Infeld would be more suitable for the position than
Mathisson who seemed to him a little ``crazy'' (AEA 37-255). It is just a few weeks
later, on 5 September 1937, that Mathisson submitted his own most important
paper on the problem of motion to the {\it Acta Physica Polonica}.

Einstein defended his assessment of Mathisson over Infeld in correspondence
with Hadamard in April 1938 (AEA 12-042, 12-043). In August 1938, he preferred Fritz London
over Mathisson, whom he could only recommend if someone else would be available to lecture
about quantum theory (HU). But later in 1938, he
recommended Mathisson again for the position, when the alternative candidate
was Reinhold F\"urth (AEA 37-271, 37-273).
As late as August 1939, Mathisson wrote to Tullio Levi-Civita from
Cambridge that he had lost his position in Poland, that Hadamard was supporting
his candidacy for the chair in Jerusalem, and that he was hoping Levi-Civita could
recommend him to Fraenkel \cite[p.~185]{Tazzioli}. By that time, however,
a decision about the position in Jerusalem had almost been reached.
It was filled in the fall
of 1939 by Giulio Racah who succeeded in the following decades to build up a
strong center for theoretical physics in Jerusalem \cite{Unna}.

\section*{1938-39: Cracow}

Some time at the end of 1937 or the beginning of 1938, Mathisson went to
Cracow at the invitation of Jan Weyssenhoff, who, in 1935, had
been  appointed  as professor of theoretical physics at the
Jagellonian  University. Unlike in Warsaw, he found there a
congenial atmosphere to work; he collaborated with Weyssenhoff,
J\'ozef K.~Luba\'nski, and  Adam Bielecki. Some time in the
1960s, Weyssenhoff told his then Ph. D. student  Andrzej Bia{\l}as that
it was Mathisson who had explained to Luba\'nski how to construct, from the
spin bivector \(s^{\mu\nu}\),  the object that is now known as the
{\it Pauli--Luba\'nski vector}.

Weyssenhoff found financial
support from private sources for Mathisson. According to
Andrzej Schinzel,  Leon Rappaport, Mathisson's
colleague from Warsaw University, was among those helping him.
Leon Rappaport wrote later a book \cite{Rapp61}; its first chapter is
devoted to Mathisson who appears under the cryptonym {\it Radon}.

A young mathematician from Cracow, Irena Jungermann, became
Mathisson's wife.

His stay in Cracow exerted a long-lasting influence on research in
theoretical physics there; Weyssenhoff and his students continued to
work on the motion of particles with structure in gravitational and
electromagnetic fields until the late 1960s. Especially important
was the extension, due to Weyssenhoff and Antoni Raabe \cite{WR}, of
Mathisson's ideas to continuous media and, in particular, the
development of a relativistic theory of ``spinning fluids''.
By integration, they obtained, from the  equations for a
dust with spin, the  Mathisson equations for individual particles.
Continuing the ideas of [M7], they found solutions of those equations corresponding to
the motion of an electron on a circle.  In another paper, they  considered the
motion of spinning particles with the speed of light \cite{WR2}.
Weyssenhoff compared the  properties of a classical
particle with spin with those described by the Dirac equation \cite{W}.
 A review
of that work is given in Ref.~\cite{BS94,Suff}; it is mentioned
in several monographs \cite{LdB}--\cite{Schm}. A recent moving homage to
Mathisson, Weyssenhoff,  Raabe and the influence of their work
is in the article by Peter Horv\'athy \cite{Horv}.

\section*{1939--40: Paris and Cambridge}

In the spring of 1939, the Mathissons went to Paris. Presumably
during the stay there, Mathisson wrote a short note on {\it Le probl\`eme
de M. Hadamard relatif \`a la diffusion des ondes\/}, that Hadamard
presented to the {\it Comptes Rendus\/} [M8]. Later that year, a longer
paper under the same title appeared in the Swedish {\it Acta
Mathematica} [M9]. It is considered to be the most
important mathematical paper by Mathisson: it presents the first proof, in
a special case, of Hadamard's conjecture on the class of hyperbolic
differential equations that satisfy Huygens' principle.

Later that year, the Mathissons went to Cambridge, where Myron
continued to work on his method of deriving equations of motion.
In a paper [M11] communicated for him by P. A. M. Dirac on 9 February 1940,
he gives a simplified derivation of his fundamental  formula \eqref{V}
and of the resulting variational principle. It is worth noting that even
though Warsaw University never offered him a job, he was loyal to it and,
even in 1940, when he could not envisage ever returning there, he indicated
that university as his affiliation. Some time in 1940, the news reached him that
Czes{\l}aw Bia{\l}obrzeski was killed in Warsaw by the Nazis. Mathisson wrote his
obituary \cite{Math40}. The tragedy concerned a different person of the same name; 
Professor Czes{\l}aw Bia{\l}obrzeski lived in Warsaw until his death in 1953.

Suffering from tuberculosis, Myron Mathisson died in England on 13~Sep\-tember~1940.

After the war, his widow, then Mrs Gill, settled in Rhodesia. In
the 1970s, Bruno Lang, a professor of radio chemistry in Warsaw,
Peter Havas and Stan Ba\.za\'nski exchanged a few letters with her.
She mentioned having seen a letter from Einstein; according to
her, he wrote ``you must be my natural son, and we must meet and talk
about this''. From those letters we know that Myron's parents came to
Warsaw from Riga, that Myron was not an easy man to reach and to
get to know, and that he died of tuberculosis, not of hunger, as suggested by
Infeld in his autobiographical sketch {\it Kordian, fizyka i ja\/}.

Mathisson had good contacts with physicists in Cambridge (DA); in the first
paper written there he thanks M. H. L. Pryce for valuable
suggestions. Mathisson made an impression on  P. A. M. Dirac, who edited and
published, posthumously, his last paper [M12] and wrote his obituary for
{\it Nature} \cite{D}.

Hadamard was so impressed by the work of Mathisson that, after his death,
living then in New York as a refugee from Nazi-occupied Paris, he published in
the prestigious {\it Annals of Mathematics\/}
a paper dedicated to Mathisson and containing an exposition of his result \cite{Had}.
It begins with the words ``To the memory of Myron Mathisson, whose premature
death is a cruel loss to Science, I dedicate this treatment of the problem
which he has solved so beautifully''.

\subsection*{Scientific publications of Myron Mathisson}

\noindent\ 1. Die Beharrungsgesetze in der allgemeinen Relativit\"atstheorie,
{\it Z.  Phys.\/} {\bf 67} (1931) 270--277.

\noindent\ 2. Die Mechanik des Materieteilchens in der allgemeinen
Relativit\"atstheorie, {\it Z.  Phys.\/} {\bf 67} (1931) 826--844.

\noindent\ 3. Bewegungsproblem der Feldphysik und Elektronenkonstanten,
 {\it Z.  Phys.\/} {\bf 69} (1931) 389--408.

\noindent\ 4.  Eine neue L\"osungsmethode f\"ur Differentialgleichungen von
normalem hyperbolischem Typus, {\it Math. Annalen\/}
{\bf 107} (1933) 400--419 and 648 (Berichtigung).

\noindent\ 5. Die Parametrixmethode in Anwendungen auf hyperbolische
Gleichungs\-systeme, {\it Prace Matematyczno-Fizyczne, Warszawa\/} {\bf 41}
(1934) 177--185.

\noindent\ 6. Neue Mechanik materieller Systeme, {\it Acta Physica Polonica\/}
{\bf 6} (1937) 163--200. An English translation by Anita Ehlers is due to
appear as a `Golden Oldie' in {\it Gen. Rel. Grav.}, forthcoming.

\noindent\ 7. Das zitternde Elektron und seine Dynamik, {\it Acta Phys. Polon.\/}
{\bf 6} (1937) 218--227.

\noindent\ 8. Le probl\`eme de M. Hadamard relatif \`a la diffusion des ondes,
{\it Comptes Rendus de l'Acad\'emie des Sciences,  Paris\/} {\bf 208} (1939)
1776--1778.

\noindent\ 9. Le probl\`eme de M. Hadamard relatif \`a la diffusion des ondes,
{\it Acta Math.\/}
{\bf 71} (1939) 249--282.

\noindent 10. Sur un th\'eor\`eme concernant une transformation d'int\'egrales quadruples
en int\'egrales curvilignes dans l'espace de Riemann,
{\it Bull. Internat. Acad. Polonaise Sci. Lett., Cracovie\/}
 Cl. Sci. Math. Natur., S\'er.  A (1939) 22--28
 (co-authors: A. Bielecki and
Jan W. Weyssenhoff).

\noindent 11. The variational equation of relativistic dynamics, {\it Proc. Cambridge
Phil. Soc.\/} {\bf 36} (1940) 331--350.

\noindent 12. Relativistic dynamics of a spinning magnetic particle,
{\it Proc. Cambridge
Phil. Soc.\/} {\bf 38} (1942) 40--60 (published posthumously).

\section*{Appendix: the sources}

Our account of Mathisson's life is based on the following sources:
Most important are several articles \cite{BS94}, \cite{BS82}
and \cite{BS92} by Bronis{\l}aw \'Sredniawa, now an Emeritus Professor
at the Jagellonian University. In 1938, as a student at that university,
\'Sredniawa met Mathisson
in person. Later, he prepared a Ph.D.~thesis, supervised by Jan
Weyssenhoff and devoted to an application of the Mathisson
variational principle to the problem of motion of dipole and
quadrupole particles \cite{BS48}. During the last 25 years \'Sredniawa
was active in describing and presenting at conferences the history
of work on relativity in Cracow. He contributed, more than
anyone else, to keeping alive the memory of Mathisson
and his collaboration with Weyssenhoff.

Another important source is the article by Peter Havas \cite{PH}.
In this historical account, Havas is very critical of the work on
the problem of motion done by Einstein and (especially) Infeld; he
gives much credit to Mathisson and writes that ``Mathisson's contributions
were far more original than Infeld's and introduced far better
mathematical methods'' \cite[p.~267]{PH}. He speculates that ``if Einstein had
succeeded in getting Mathisson to join him in Berlin or
Princeton ... there would have been no EIH, but presumably an EMH ...'' \cite[p.~268]{PH}.
Infeld himself also mentions Mathisson in his
autobiography \cite[203--205]{Infeld1967}.

Among the unpublished sources, the most important one is the
Einstein--Mathisson correspondence as well as related documents in
the Albert Einstein Archives (AEA) at the Hebrew University of Jerusalem,
see note~\ref{noteAEA}. Unpublished documents from the Albert Einstein Archives
are quoted by kind permission. Further documents, including a CV dated 1937, 
are contained in the Hebrew University's archives (HU), 
folder `physics 1938'; we wish to thank
Issachar Unna for his assistance in locating these documents. 
We are also grateful to { Stanis{\l}aw
Ba\.za\'nski} who made available to us his excerpts from the
University of Warsaw Archive; they are referred here to as (UW).
He also supplied us with two letters from Mrs Irena
Gill (IG), Mathisson's widow, the portrait of Mathisson reproduced as Fig.~1,
and a letter, dated 30 October 1978, from
Prof.~B.~L.~Laptev, University of Kazan (BLL). We also thank
W{\l}odzimierz Zych who helped us to obtain information about
Mathisson's studies at Politechnika Warszawska (PW).

The Rockefeller Archive Center in New York (RA) contains documents pertaining
to Einstein's visit and intervention about Mathisson (RFam, RG2, Ser.~H, Box 162, Folder 1256).
The Paul Dirac papers at Florida State University in Tallahassee (DA) contain two
letters by Mathisson and one by his widow, dated 5 February, 7 April, 22 November 
1940, respectively, as well as a letter by
Hadamard to Dirac, dated 14 November 1940. The Niels Bohr Archives (NBA) in Copenhagen 
contain a letter by Mathisson, dated 7 August 1937, as well as a response by 
Bohr (Suppl. Sci. Corr.).
Finally, three letters by Mathisson to Levi-Civita, dated 4 November 1932,
26 September 1937, and 25 August 1939,
as well as two letters by Hadamard to Mathisson, dated 4 November 1931 and 
9 February 1932, were published in \cite{Tazzioli}. We also thank Diana Buchwald,
Margaret Hogan, Rosy Meiron, Felicity Pors, Sharon Schwerzel, and Rossana Tazzioli 
for further information and assistance.

\end{document}